\documentclass[sigconf]{acmart}

\AtBeginDocument{%
  \providecommand\BibTeX{{%
    \normalfont B\kern-0.5em{\scshape i\kern-0.25em b}\kern-0.8em\TeX}}}

\setcopyright{acmlicensed}

\copyrightyear{2024}
\acmYear{2024}
\setcopyright{rightsretained}
\acmConference[GoodIT '24]{International Conference on Information
Technology for Social Good}{September 4--6, 2024}{Bremen, Germany}
\acmBooktitle{International Conference on Information Technology for Social
Good (GoodIT '24), September 4--6, 2024, Bremen,
Germany}\acmDOI{10.1145/3677525.3678685}
\acmISBN{979-8-4007-1094-0/24/09}

\usepackage{comment}

\begin{document}

\title[Mapping Child Malnutrition through Location-based Games]
{Mapping Child Malnutrition and Measuring Efficiency of Community Healthcare Workers through Location Based Games in India}

\author{Arka Majhi}
\email{arka.majhi@iitb.ac.in}
\affiliation{
  \institution{Indian Institute of Technology Bombay}
  \streetaddress{Powai}
  \city{Mumbai}
  \state{Maharashtra}
  \country{India}
  \postcode{400076}
}

\author{Satish B. Agnihotri}
\email{sbagnihotri@iitb.ac.in}
\affiliation{
  \institution{Indian Institute of Technology Bombay}
  \streetaddress{Powai}
  \city{Mumbai}
  \state{Maharashtra}
  \country{India}
  \postcode{400076}
}

\author{Aparajita Mondal}
\email{aparajita.mondal@tuni.fi}
\affiliation{
  \institution{Tampere University}
  \streetaddress{Kalevantie 4}
  \city{Tampere}
  \state{}
  \country{Finland}
  \postcode{33100}
}

\renewcommand{\shortauthors}{A. Majhi, et al.}

\begin{abstract}

In India, Community Healthcare Workers (CHWs) serve as critical intermediaries between the state and beneficiaries, including pregnant mothers and children. Effective planning and prioritization of care and services necessitate the collection of accurate health data from the community. Crowdsourcing child anthropometric data through CHWs could establish a valuable repository for evidence-based decision-making and service planning. However, existing platforms often fail to maintain CHWs' engagement over time and across different spatial contexts, resulting in spatially misrepresented and outdated data.

This study addresses these challenges by conducting a co-design exercise to develop innovative methods for collecting anthropometric data over time and space. The exercise involved analyzing data to create hotspot and density distribution maps. We implemented a trial of the developed game with two groups (n=94 per group) from various states across India, comparing the game-based and non-game-based data collection methods. Our findings reveal that the game-based approach significantly improved measuring efficiency (p<0.05) and demonstrated superior engagement and retention compared to the non-game-based method.

This research contributes to the expanding literature on co-design and Research through Design (RtD) methodologies for developing geospatial games, highlighting their potential to enhance data collection practices and improve engagement among CHWs.

\end{abstract}



\begin{teaserfigure}
  \includegraphics[width=0.225\paperwidth]{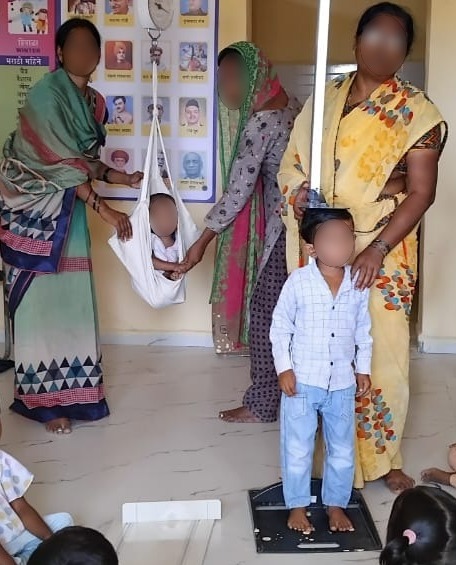}
  \includegraphics[width=0.31\paperwidth]{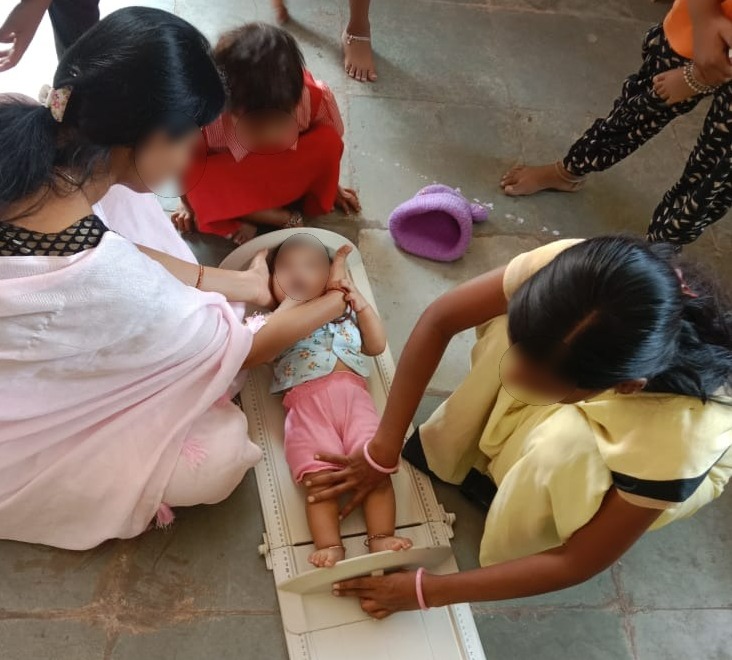}
  \includegraphics[width=0.275\paperwidth]{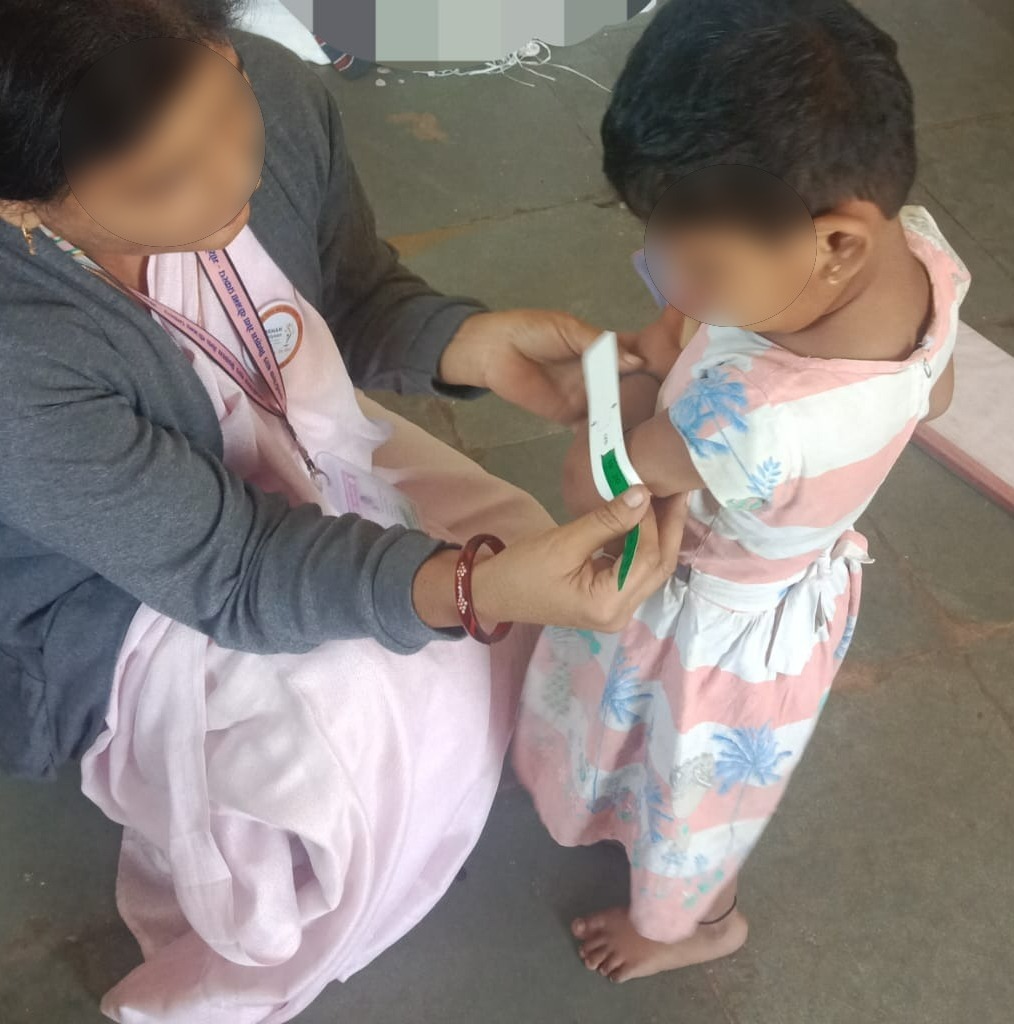}
  \caption{Left: Measuring height of a child by making him stand against a stadiometer and weight of another child by putting her in a pouch, hung from a salter scale and supported by the CHW and her mother for safety concerns ; Middle: Measuring height of  neonatal by resting her on an infantometer ; Right: Measuring Mid Upper Arm Circumference (MUAC) of a child on her left arm by using color coded MUAC tape}
  \label{fig:Measurement}
\end{teaserfigure}

\maketitle

\section{Introduction}

The early identification and management of child undernutrition is crucial for improving health outcomes and saving lives \cite{Heidkamp2021}. In India, the latest National Family Health Survey (NFHS-5, 2019–2021) \cite{NFHS5IndiaFactsheet} provides population-based anthropometric data for children under five years of age. However, these surveys are conducted with a substantial gap of nearly five years, making it challenging to maintain accurate and up-to-date geospatial representations of health issues within communities, and the access to health infrastructure available. Continuous data collection is essential during the first 1000 days of a child's life to ensure effective care and service planning.

Community Health Workers (CHWs) are instrumental in enhancing health outcomes in resource-limited settings by providing education, vaccinations, and health monitoring services. The widespread use of smartphones among CHWs presents a significant opportunity to enhance nutrition surveillance. Technology-enabled solutions can reduce costs and increase the reach and efficiency of nutrition information systems, thereby revolutionizing data-driven decision-making. Moreover, these technologies can improve the accuracy of nutrition monitoring by automating complex calculations, such as finding z-scores from WHO Growth Tables.

India's 'Poshan Tracker' is the largest global nutrition surveillance system operated via mobile phones. It offers transparent data on anthropometric outcomes, the functioning of Anganwadi Centres (AWCs), and the delivery of care services, including supplementary food for women, children, and adolescent girls. Despite these efforts, measuring efficiency—encompassing accuracy, speed, and overall effectiveness—remains inadequate. Challenges such as insufficient training, supervision, and resources, coupled with limited literacy and smartphone skills, hinder CHWs' performance. Many CHWs continue to use manual registers for recording anthropometric measurements. Ensuring high-quality work by CHWs requires ongoing training and support \cite{Mishra2024}. Discussions with Women in Global Health (WGH) India and Accredited Social Health Activists (ASHA) workers highlight the need for strategies to enhance capacity through technology and improved supervision \cite{Asthana2022}.

Games and playful activities offer a promising framework for crowdsourcing health data in the context of citizen science. Engaging gameplay has been shown to enhance motivation and learning \cite{Bochennek2007}. This study aims to compare the effectiveness of game-based versus traditional data entry methods for crowdsourcing health data. The involvement of CHWs in developing these applications has often been neglected, leading to lower adoption of digital tools and resistance from CHW communities. Protests by CHWs against the adoption of the Poshan Tracker app have been reported across various regions in India. These protests reflect concerns over the system's implementation, usability, and impact on their daily work. CHWs have raised issues related to the system's complexity, inadequate training, and the perceived increase in their workload. The resistance highlights the need for a more user-centered design and supportive implementation strategies to ensure effective adoption and utilization of digital tools in health monitoring. Despite progress in app development, there remains a limited understanding of the challenges faced by emerging smartphone users in resource-constrained environments. This research employs a multidisciplinary approach, combining qualitative feedback from CHWs with gameplay analytics to address these issues.

The research questions this study tries to answer are:
\begin{itemize}
    \item [\bfseries RQ1] How does measuring efficiency differ for participants using a traditional non-game data entry app compared to a game-based app?
    \item [\bfseries RQ2] Is the retention of measuring efficiency similar in both methods?
\end{itemize}

This research makes several significant contributions:
\begin{itemize}
    \item [\bfseries Literature] It broadens the literature by investigating how CHWs conceptualize the integration of child anthropometric measurement into geospatial game concepts, employing a Research through Design (RtD) approach \cite{Zimmerman2007}.
    \item [\bfseries Concept] It introduces a novel game concept designed to engage CHWs in the collection of anthropometric data for children.
    \item [\bfseries Tool] It presents an Android game developed using Unity, available for free on Google Play Store and Apple App Store.
    \item [\bfseries Repository] It provides an open-access Unity project, including code and materials, on GitHub for further modification and development.
\end{itemize}

\subsection{Background and Related Works}

Previous research has explored various methods of integrating games and playful activities into the training and professional development of Community Health Workers (CHWs) via smartphone applications. Researchers \cite{Majhi2021} investigated the effectiveness of quiz-based refresher training delivered through smartphones. Building on this, researchers \cite{Shah2017} enhanced the quiz format by incorporating video clips and incentivizing correct responses with talk time for cellular networks. Further advancements were made \cite{Majhi2022}, by exploring Augmented Reality (AR) games for collaborative play, and examining the potential of serious card games for refresher training of CHWs \cite{Majhi2024a,Majhi2024c}.

Despite these innovations, none of the prior studies have investigated the use of games for crowdsourcing anthropometric data of children. Integrating such games into citizen science initiatives could significantly contribute to evidence-based decision-making for community welfare and planning. By leveraging game-based approaches, researchers and practitioners could enhance data collection efforts, ultimately supporting better health outcomes and resource allocation.

\section{Method}

This research investigates the effectiveness of a game-based tool for anthropometric measurement designed specifically for CHWs. It contributes to the expanding literature on digital tools for healthcare data collection, with a focus on the use of smartphones in resource-constrained settings, particularly in India.

\subsection{Aim}

The primary aim of this study is to assess the efficacy of digital crowdsourcing tools, particularly data entry games, in enhancing the measurement efficiency of CHWs operating in resource-limited environments. This research seeks to provide insights into the design, implementation, and evaluation of such measurement efficiency games. Additionally, it aims to explore the experiences of CHWs during gameplay and assess the impact of the game on their daily work routines.

Involving CHWs in the design and development process ensures that the game is customized to their needs and preferences, thereby increasing its relevance and usability. Qualitative feedback from CHWs is crucial to create a user-friendly and engaging tool that aligns with their practical requirements. This research not only advances the literature on digital tools for healthcare training but also supports the growing evidence base for using co-design approaches in game development.

\subsection{Participants}

G-power \cite{Erdfelder2007} was used to calculate the sample size. The effect size convention (d)=0.5, power (1-\(\beta\))=0.95, and level of significance (\(\alpha\))=0.05 was chosen. The calculated sample size is 88 for each arm. Considering the 10\% attrition rate, 97 to 100 participants were required for each groups. CHW supervisors were contacted, and 200 CHWs were selected through convenience sampling. Till the end of the study, 94 participants retained from each group. The attrition was largely due to the unavailability of the CHWs on the days of test and evaluation due to emergency visits. For some CHWs their other partnering group member was missing, thus needed to be excluded from the study. The demographic details of the participants are given in Table \ref{tab:ParticipantsDemography}.

The study involves two groups for a quasi-experiment. 
Firstly ones measuring and entering anthropometric data through the game app  (n=94) (IG),
secondly ones measuring and entering anthropometric data through the regular data entry app  (n=94) (CG),
Each group consists of 94 CHWs. 

Each intervention group is trained to use the game app and the measuring efficiency is compared between them.

\begin{table*}  
  \begin{tabular}{l cc}
 \toprule 
 Groups & Control Group (n=94) & Intervention Group (n=94)\\
 \cmidrule(lr){2-3}
 Parameters & Using Regular Monitoring App & Using Game based Monitoring App\\
 \midrule
 Age &  \\
 Less than 30               & 2 (2.13\%)  & 4 (4.26\%)  \\
30-40                       & 46 (48.94\%) & 48 (51.06\%) \\
40-50                       & 42 (44.68\%) & 41 (43.62\%) \\
No information              & 4 (4.26\%)  & 1 (1.06\%)  \\
 \hline
 Education (grade) &  \\
$Below  8^{th}$             & 2 (2.13\%)  & 2 (2.13\%)  \\
$8^{th}-10^{th}$            & 18 (19.15\%) & 20 (21.28\%) \\
$10^{th}-12^{th}$           & 64 (68.09\%) & 62 (65.96\%) \\
Graduate and above          & 8 (8.51\%)  & 10 (10.64\%) \\
No information              & 2 (2.13\%)  & 0 (0.00\%)  \\
 \hline
 Experience as CHWs (years) &  \\
0-5                         & 8 (8.51\%)  & 8 (8.51\%)  \\
5-10                        & 34 (36.17\%) & 38 (40.43\%) \\
10-15                       & 38 (40.43\%) & 36 (38.30\%) \\
Above 15                    & 14 (14.89\%) & 10 (10.64\%) \\
No information              & 0 (0.00\%)  & 2 (2.13\%)  \\ 
 \bottomrule
 \\
 \end{tabular}
 \caption{Participants' Demography}
  \label{tab:ParticipantsDemography}
 \end{table*}

  \subsection{Ethical Considerations}
The ethical conduct of this study was approved by the Institute Review Board at the Indian Institute of Technology Bombay, India (Approval No: IITB-IRB/2022/051), in accordance with the guidelines set forth in the Declaration of Helsinki (2013) \cite{DOH2013}. Prior to participation, all participants were verbally informed about the objectives and procedures of the study. Written consent was obtained from each participant before administering any surveys, and this consent was documented within the questionnaire. The findings of the study, including survey results, were shared with the supervisors of the Community Healthcare Workers (CHWs). To ensure confidentiality, all participant names and other identifying information were anonymized in the study reports.

\subsection{Statement of Positionality}

All three authors have extensive experience conducting fieldwork with Community Health Workers (CHWs) in rural India, with a particular focus on child and maternal health. Together, the authors strive to represent the perspectives of CHWs in marginalized and low-resource contexts as accurately as possible.

\subsection{Designing the Game Concept - Framing Challenges and Rules}

The primary objective of the game is to facilitate Community Health Workers (CHWs) in measuring and entering anthropometric data of children into a crowdsourced, game-based platform. Given that CHWs are predominantly middle-aged or elderly women, the design of the game needed to account for the unique needs and preferences of this demographic. Research has shown that personalizing gaming experiences is crucial for older adults, who prioritize enjoyment over performance \cite{Kappen2016, Birk2017}. Therefore, addressing age-specific challenges and barriers was essential in ensuring the relevance and effectiveness of the game design.

The co-design exercise with CHWs, comprising 20 participants organized into 10 pairs, involved a structured six-stage process:

\begin{itemize}
\item \textbf{Building Bridges:} Facilitators initiated the process by explaining the importance of periodically collecting crowdsourced anthropometric data of children within the community. This stage aimed to establish a clear understanding of the project's objectives among participants.

\item \textbf{Developing a User Model:} Participants clarified their understanding of the serious game's goals. This stage helped ensure that the game design aligned with the intended outcomes and addressed the needs of CHWs effectively.

\item \textbf{Mapping Possibilities:} Participants engaged in discussions with their group members to exchange and refine ideas. They were encouraged to enact the game, articulating their decisions and actions as if they were playing in real life. This interactive approach helped in visualizing and iterating the game mechanics.

\item \textbf{Developing Prototypes:} Groups were tasked with drawing out the steps or stages of the game they had conceptualized. This stage focused on translating ideas into tangible prototypes, allowing for a more concrete representation of the game design.

\item \textbf{Eliciting Feedback:} Each group presented their prototype to the entire team and received feedback from peers. This collaborative feedback process facilitated the refinement of game concepts and the identification of potential improvements.

\item \textbf{Implementing Changes:} The iterative process of refining the game design continued with modifications based on feedback. Participants explored how changes to specific game mechanics could impact the overall gameplay experience. This stage ensured that the game design evolved to meet the practical needs and preferences of the CHWs.

\end{itemize}

This iterative and participatory design process enabled the development of a game that was both engaging and practical for the target user group, integrating their insights and feedback into the final game concept. The co-design exercise revealed significant thematic variations in the ideas generated. 

However, some common game mechanics emerged that are particularly effective in promoting geospatial measurement and data collection. These mechanics predominantly focused on encouraging participation either across spatial dimensions or time.

\begin{itemize}
    \item Sustained Participation over Spatial dimensions: Leveraging gameful strategies can be effective in motivating participants to explore various areas and gather valuable data. One approach could be the use of incentivized exploration games, which reward participants for collecting data in previously uncharted regions \cite{JuhoHamari2019}. These rewards can take many forms, such as virtual badges, in-game currency, or real-world incentives, all of which can significantly boost engagement.
    
    The design of game locations is crucial in influencing players' socio-spatial behavior and perceptions of mobility within the gaming environment. Platforms like Foursquare \cite{Jordan2013} have demonstrated how game elements can transform ordinary locations into sites of adventure and discovery. By turning the mundane into the extraordinary, these platforms encourage users to explore areas they might otherwise ignore. In this context, we can make remote or hard to reach areas in the community (due to geographical or access problem) more attractive to the CHWs to visit and collect data. By making data collection in underserved areas a compelling activity, could potentially gather valuable information that can be used to improve infrastructure and services in these regions. This strategy transforms data collection into an enjoyable and rewarding experience, blending the excitement of discovery with the practical benefits of collected rich and granular community healthcare data.

    \item Sustained Participation over Time: Sustained participation over time is critical for the long-term success of gamified initiatives. To achieve this, various strategies can be employed. Seasonal campaigns and events are particularly effective, as they refresh content periodically and provide participants with new reasons to engage. These campaigns can be themed around different times of the year or significant events, ensuring that there is always something new and exciting for participants to look forward to.
    
    Using progressive storylines and narratives is another powerful strategy for maintaining engagement. By developing ongoing stories that unfold over time, we can keep participants invested in the outcome and eager to see what happens next. This approach is particularly effective for community healthcare monitoring, where continuous engagement is essential for gathering comprehensive data.
    
    Providing participants with updates on the real-world impact of their contributions is also crucial. When participants understand how their efforts are making a tangible difference, they are more likely to stay engaged. For example, if participants are collecting data on child anthropometry, updates could highlight how their contributions are helping to identify underserved areas and improve healthcare. This not only validates their efforts but also reinforces the importance of their participation.
    
    By linking game activities with meaningful outcomes, we can ensure sustainable engagement. When participants see that their actions have a real-world impact, they are more likely to remain committed over the long term. This approach not only fosters long-term participation but also builds a sense of community and shared purpose among participants.

\end{itemize}

The co-design exercise revealed a diverse array of game themes, highlighting the potential for a flexible gaming framework with common mechanics, enabling communities to develop their thematic variations. This approach not only fosters creativity but also enhances engagement by allowing participants to customize their gaming experiences. However, to validate the concept of crowdsourcing data collection through a collaborative game, we developed a basic game application for field testing.

Adopting a techno-optimistic mindset, we refrained from rejecting any game ideas outright. Instead, we explored a wide range of potential and feasible concepts. This open-minded approach facilitated further ideation in group settings, leading to the development of a prototype game that incorporated essential mechanics to ensure both engagement and enjoyment. The common game mechanics identified during the co-design exercise primarily aimed at encouraging geospatial measurement and data collection by promoting participation across spatial dimensions or temporal periods. For instance, spatial participation was encouraged through exploration games that rewarded players for collecting data in various locations, while temporal participation was sustained through progressive storylines, seasonal campaigns, and real-world impact updates.

Field testing was a critical component of our research methodology. We followed participants in their workspaces, instructing them to simulate playful activities as if they were engaged in the actual game. This approach provided valuable insights into their interactions with the game mechanics and their overall playtesting experience. Observing gameplay in real-world contexts allowed us to identify potential issues and areas for improvement, ensuring that the final game design would be both practical and engaging.

Several key insights emerged from the development and testing process. Firstly, allowing communities to create their thematic variations enhances engagement and fosters a sense of ownership and creativity. This customization can lead to higher levels of participation and more diverse data collection. Secondly, incentivizing exploration and data collection through rewards and recognition proved to be an effective strategy. Tailoring these rewards to the specific interests and motivations of the target audience further enhanced engagement. Thirdly, progressive storylines and real-world impact updates are crucial for maintaining long-term participation. Participants are more likely to remain involved when they see tangible benefits from their contributions and feel connected to an ongoing narrative. Finally, simulating gameplay in participants' actual work environments provided valuable practical insights, helping to refine the game design to ensure it was both effective and enjoyable.

\begin{figure*}
    \includegraphics[width=0.3\paperwidth]{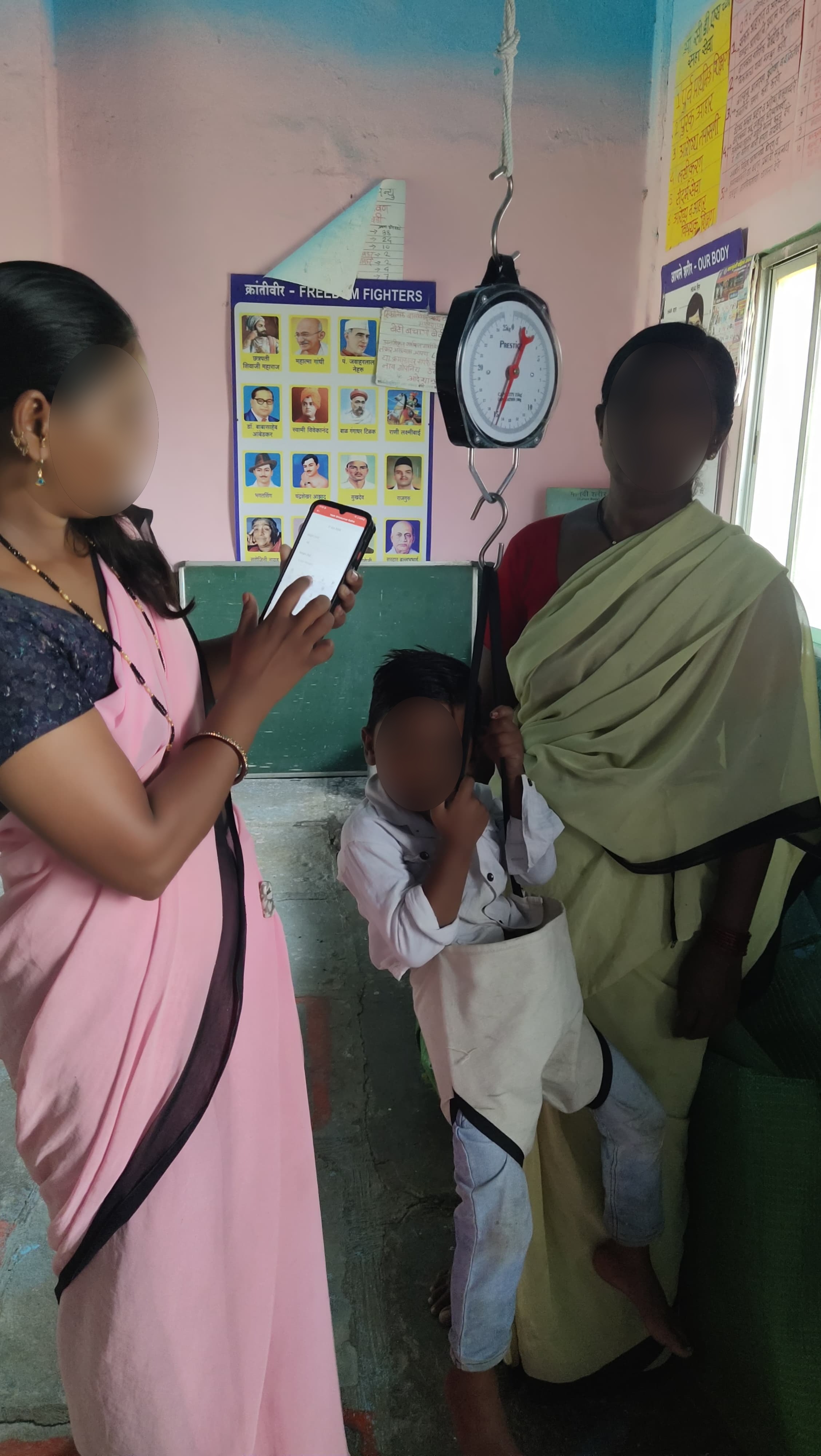}
  \includegraphics[width=0.4\paperwidth]{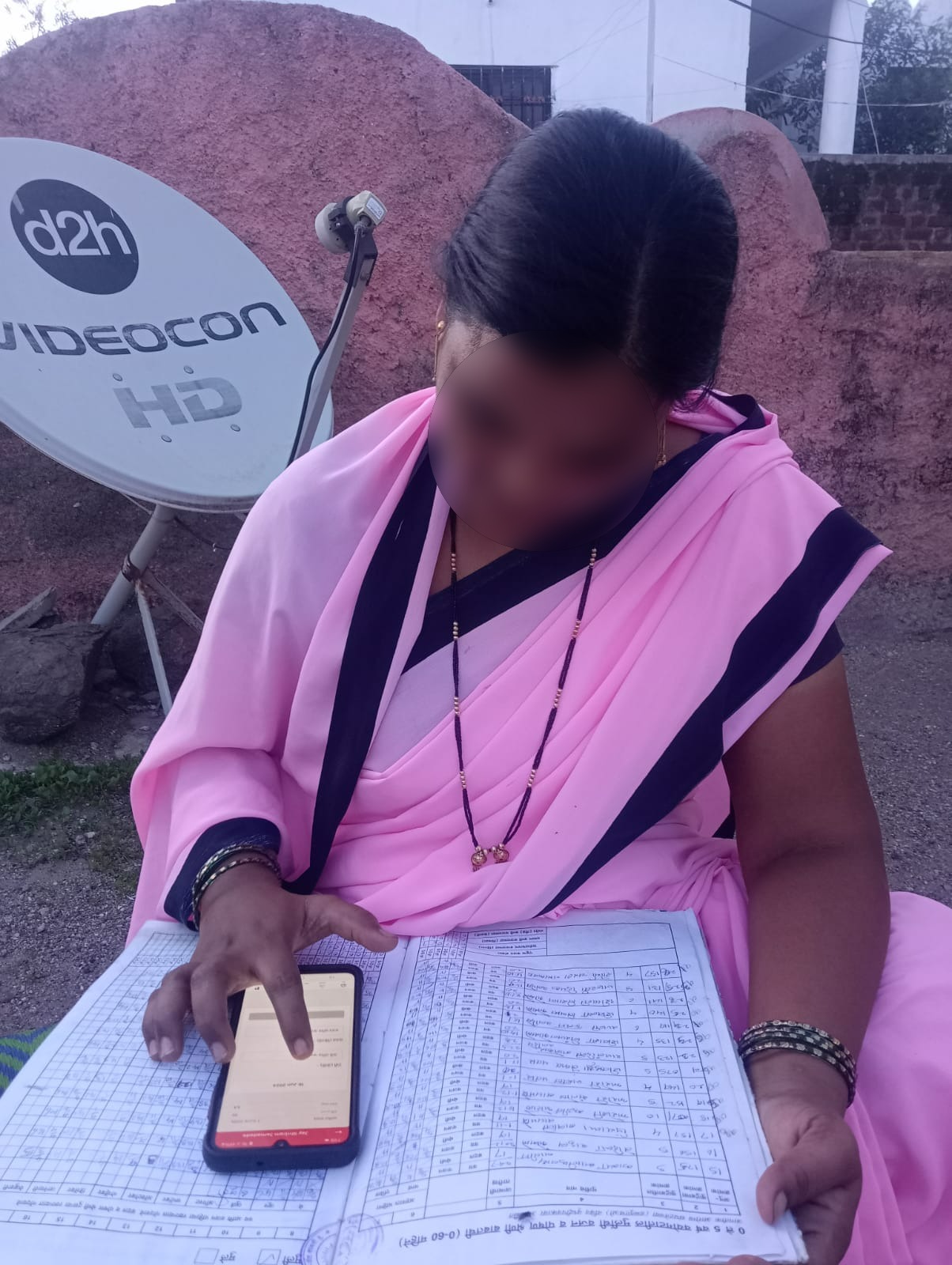}  
  \caption{Left: Entering data into smartphone after watching the weight measurement in the salter scale ; Right: Entering data into physical registers and copying it into smartphone}
  \label{fig:Data_Entry}
\end{figure*}

Figure \ref{fig:Data_Entry} shows how a CHW measures and record the data into the game based platform. The figure also shows how they used to do it previously where one had to enter the data manually into the registers and enter them into their smart phone app in order to digitize them and sync to the server.

\begin{figure*}
    \includegraphics[width=0.8\paperwidth]{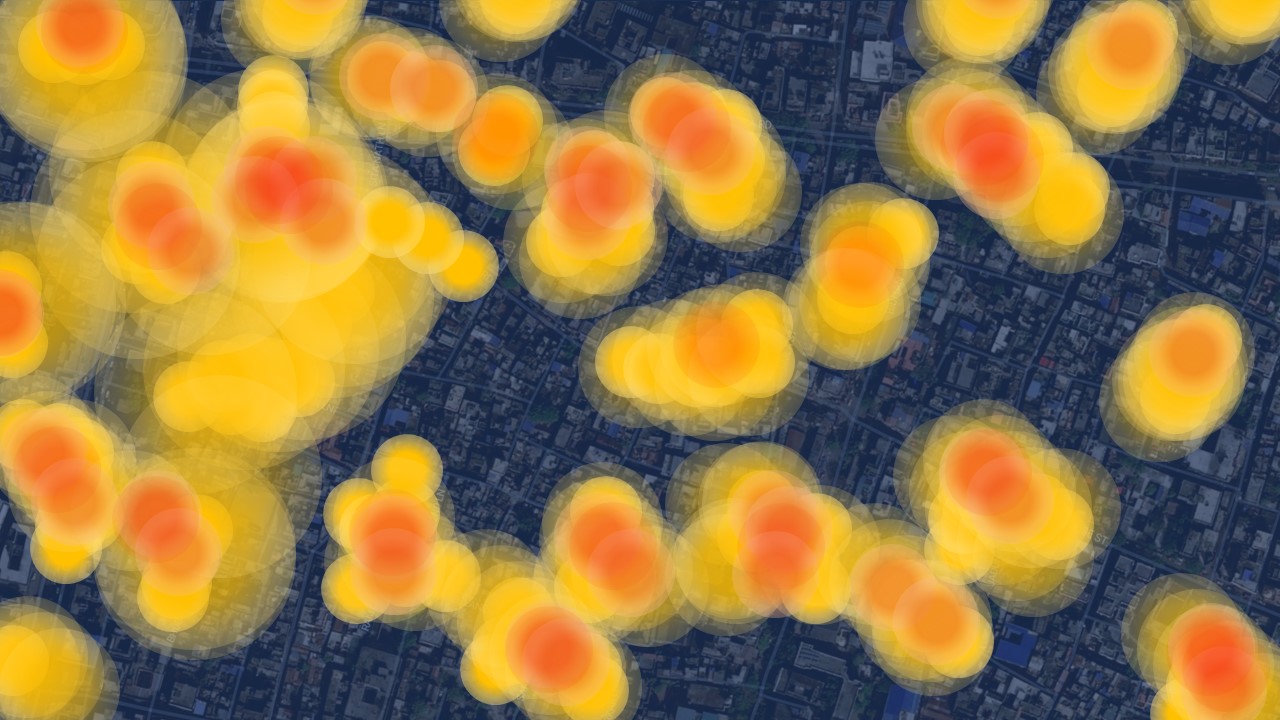} 
  \caption{Hotspot Distribution map or Density Map of prevalence of child malnutrition mapped through data crowdsourced by CHWs through playing the game}
  \label{fig:Hotspot_Distribution}
\end{figure*}

After analysis of the data, a hotspot distribution is formed for each measurement efficiency as a factor. Fig \ref{fig:Hotspot_Distribution} shows a reference hotspot distribution of child malnutrition in a city precinct created through the crowdsourcing efforts of CHWs while playing the game.

\subsection{Experimental Design}
A quasi-experimental design (QED) was selected over a true experimental design (TED) for this study due to practical constraints. Conducting a TED was not feasible in this context because of the shared geolocation and relatively small sample size, which made randomization challenging \cite{Harris2006}. Although QEDs generally have lower internal validity compared to TEDs, they offer higher external validity since the intervention is followed by the measurement of outcomes \cite{Harris2006}. In this QED, the game's design serves as the primary method for evaluating the benefits of the intervention. This approach allows for the practical assessment of the game's impact in a real-world setting, providing valuable insights into its effectiveness.

\subsection{Evaluation}
The baseline data for the last month was collected from secondary sources to assess the existing measurement efficiency of the Community Health Workers (CHWs). Following this, the intervention was conducted with the intervention group (IG). To account for placebo effects, the control group (CG) was given similar attention and focus without the actual intervention of the game-based platform.

After one month of intervention, a post-test was administered to evaluate any changes in the measurement efficiency of the CHWs. It is important to note the potential for a novelty effect, where initial adoption shows significant improvement, which may diminish over time. Therefore, understanding the long-term effects of the intervention is crucial, necessitating longitudinal research.

To address this, a delayed retention test (DRT) was conducted three months post-intervention, mirroring the post-test conditions. The players were not briefed or informed about the DRT in advance, ensuring an unbiased assessment of the intervention's lasting impact. Throughout the study, the CG did not receive any intervention. However, after the study concluded, the CG was trained to match the benefits received by the IG, thereby bridging any knowledge and training gaps. This step ensures ethical considerations and the equitable distribution of training benefits across all participants.

\section{Findings}

\begin{table*}  
  \begin{tabular}{l cc cc}
 \toprule
 & \multicolumn{2}{c}{Control Group (CG) (n=94)} & \multicolumn{2}{c}{Intervention Group (IG) (n=94)} \\
  & \multicolumn{2}{c}{Using Regular Monitoring App} & \multicolumn{2}{c}{Using Game based Monitoring App} \\
 \cmidrule(lr){2-3} \cmidrule(lr){4-5}
 ~ & Mean & Median & Mean & Median\\
 Test phase & (SD) & (min-max)  & (SD) & (min-max) \\
 \cmidrule(lr){2-3} \cmidrule(lr){4-5}
Baseline & 51.46 (9.21) & 49.85 (36-70) & 49.04 (10.57) & 48.8 (18-72) \\
Post-test & 54.84 (14.96) & 53.84 (29-84) & 73.9 (14.28) & 75.82 (46-111) \\
Long Post-test & 52.58 (13.59) & 52.6 (28-79) & 69.14 (16.63) & 70.79 (26-108) \\
  (After 3 months)\\ 
 \bottomrule
 \\
 \end{tabular}
 \caption{Test Scores}
  \label{tab:TestScores}
\end{table*}

\begin{figure*}
    \includegraphics[width=0.75\paperwidth]{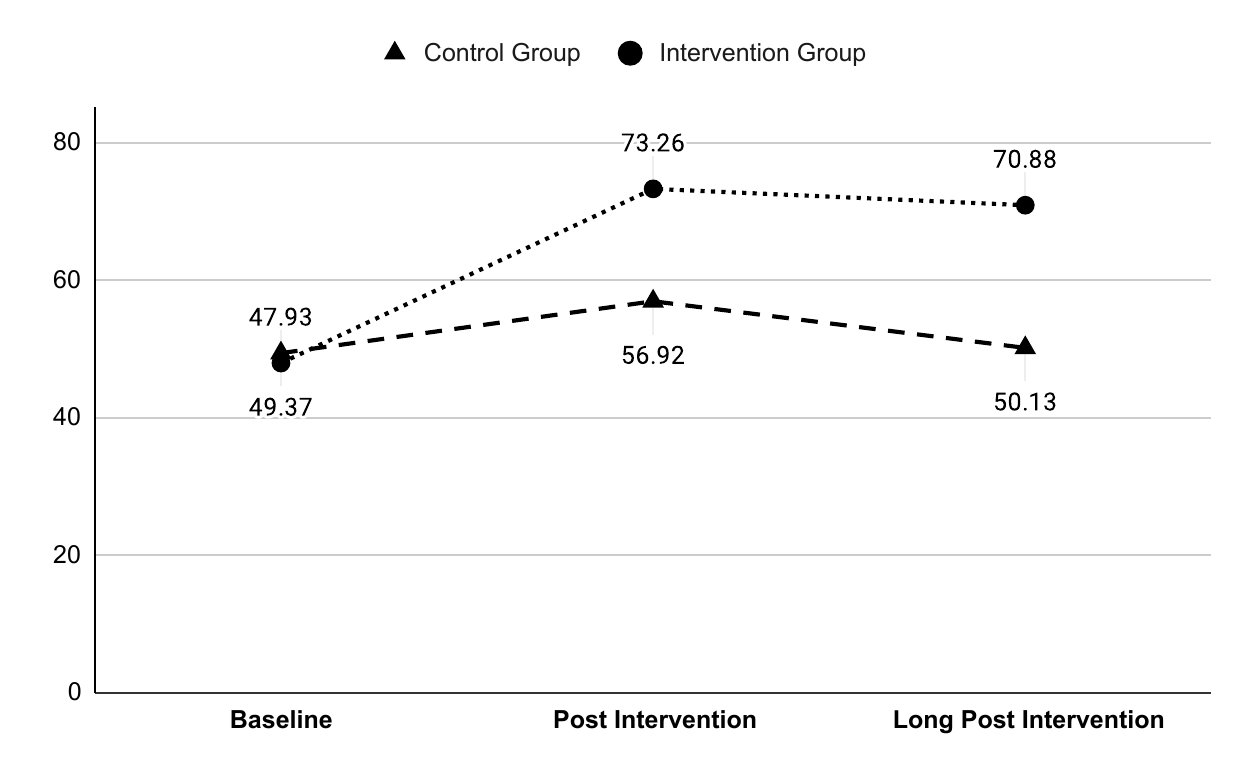}
  \caption{Line chart showing trends of change of mean value in baseline, post-test and long-post test across the control and the intervention group}
  \label{fig:Chart}
\end{figure*}

Table \ref{tab:TestScores} tabulates the scores of both groups. After testing assumptions for t-tests like normality and sphericity for both group scores, Pairwise t-tests were performed to check within-group and between-group differences at every phase of the evaluation.

The Intervention Group (IG) performed significantly better than the Control Group (CG) (p = 0.00004; p < 0.05) with a substantial effect size (Cohen's D = 1.6), demonstrating the effectiveness of the game-based application over the simple data entry procedure, addressing Research Question 1. However, long-term post-test trends for both groups showed a similar decline in scores, indicating that interest in the game-based data entry procedure may diminish over time without the reinforcement of game incentives. Despite this decline, the game-based approach still outperformed the non-game and non-incentivized data entry procedure, addressing Research Question 2. In summary, the game application significantly improved measurement efficiency and retention of skills after three months.

\section{Discussion}

\subsubsection{Trends:}

The analysis revealed that years of formal education did not have a significant effect on the baseline scores of Community Health Workers (CHWs). This finding suggests that the formal education received by CHWs does not directly correlate with their initial measurement efficiency in the baseline assessment. Instead, the knowledge and proficiency exhibited by CHWs are more strongly influenced by their practical experience and hands-on involvement in the field.

The disparity between formal education and practical effectiveness can be attributed to the nature of the CHW role, which often involves dealing with a wide range of real-world issues that may not be fully addressed through formal educational programs. The years of experience accumulated while working in the field, coupled with direct engagement in addressing common health challenges, play a crucial role in shaping the skills and effectiveness of CHWs. This practical experience equips CHWs with contextual knowledge and problem-solving abilities that are not always captured through formal academic training.

Furthermore, the emphasis on hands-on experience highlights the importance of fieldwork in developing competencies relevant to the role of CHWs. While formal education provides a foundational understanding of health principles and practices, it is the direct application of this knowledge in real-world scenarios that significantly enhances the effectiveness of CHWs. This practical approach to learning and skill development is essential for addressing the complex and dynamic nature of community health work.

This insight emphasizes the need for training programs and interventions to focus on experiential learning and field-based practice to improve the effectiveness of CHWs in their roles.

\subsubsection{Malpractices:}
During the study, random checks revealed instances of malpractice among some Community Health Workers (CHWs), specifically data falsification. These individuals were excluded from both the study and the statistical evaluations to ensure the integrity of the data. The observed malpractice highlights a potential incentive for CHWs to falsify data, as it can create a misleadingly positive representation of the state of healthcare within the community.

This issue underscores the necessity of implementing robust checks and balances within the application to detect and flag instances of data manipulation. Incorporating such mechanisms is essential to maintain the credibility and reliability of the data collected. Potential solutions could include automated alerts for unusual data patterns, regular audits, and validation procedures to ensure data accuracy. Addressing these concerns proactively will enhance the effectiveness of the app and ensure that the data used for evaluating healthcare interventions is both accurate and reliable.

\subsubsection{Play-testing experiences}
After engaging with the game for 2-3 rounds, Community Health Workers (CHWs) typically became familiar with the sequences and required less assistance. This familiarity led to a notable shift in behavior: CHWs began to form teams and actively sought enjoyment through competitive play, often challenging their opponents. This camaraderie and competitive spirit enhanced their engagement with the game.

However, observations during 2-3 weeks of shadowing revealed some potential issues. Occasionally, teams would criticize the knowledge of their opponents during gameplay. While this competitive edge can increase the challenge and excitement for players, it also has the potential to demotivate opponents. Such negative interactions could undermine the collaborative nature of crowdsourcing efforts and impact the overall effectiveness of the game.

To address these concerns, it is important to incorporate features into the game that foster a positive and supportive environment. Implementing mechanisms to moderate or manage competitive behavior and ensure respectful interactions can help maintain motivation and engagement among all participants. These adjustments will be crucial in balancing the competitive elements of the game with its collaborative goals, ultimately enhancing both the gameplay experience and the quality of data collected.

\subsubsection{Real-life Outcomes}

During the design and evaluation phase of the game application, we conducted shadowing of the Community Health Workers (CHWs) to assess the impact of the app on their motivation and the quality of their daily activities. Our observations indicated that while the game app positively influenced CHWs' engagement and enthusiasm in their tasks, this effect was predominantly short-term.

These findings are consistent with existing literature on gamification and its effects on work-related behavior. Research \cite{Deterding2018} suggests that while gameplay can enhance engagement and motivation in the short term, it may not necessarily lead to sustained behavior change or long-term improvements in job outcomes. The initial enthusiasm and increased motivation observed among CHWs following their engagement with the game app align with this perspective, reflecting the potential of gamification to temporarily boost interest and participation.

However, the lack of evidence for long-term behavior change or sustained improvements in job performance raises important considerations. The temporary nature of the engagement suggests that without ongoing reinforcement or integration of game mechanics into daily routines, the positive effects of the game app may diminish over time. This highlights the need for continuous support and possibly the incorporation of adaptive features within the game to maintain long-term engagement and effectiveness.

\subsubsection{Limitations of the study}
The experiment was conducted within formal institutional settings, where the Community Health Workers (CHWs) operate under a structured top-down bureaucracy. This institutional framework may have constrained the extent to which CHWs could engage in collaborative gameplay or provide candid feedback to the researchers. The hierarchical nature of the system might influence their interactions and limit their willingness to freely share insights or critique, thereby impacting the effectiveness of team-based gaming interventions.

Despite the limited number of participants in the co-design exercise, the participants were ethnically diverse, representing various communities. This diversity ensured that the game design concepts were reflective of the broader CHW population. The repeated themes, game mechanics, and design ideas that emerged from the exercise suggest that the game has the potential to engage a wide range of CHWs. These insights indicate that location-based games could be a valuable tool for crowdsourcing child anthropometric data collection, offering a dynamic and interactive approach to data gathering.

\section{Novelty} 

To the best of our knowledge, no previous research studies have utilized a serious game approach to map malnutrition in the community through the crowdsourcing efforts of Community Health Workers (CHWs). This novel approach leverages games to engage CHWs in data collection activities, providing a fresh perspective on how technology and game-based strategies can be employed to address pressing public health issues such as malnutrition. By integrating serious games with crowdsourcing, this research explores an innovative method to enhance the efficiency and effectiveness of community-based health data collection initiatives. 

\section{Conclusion}

The study demonstrates the efficacy of using a game-based app for collecting child anthropometric data through the crowdsourcing efforts of Community Health Workers (CHWs). The co-design of geospatial games for collecting child anthropometric data notably increased the motivation of Community Health Workers (CHWs), leading to more frequent home visits and enhanced data collection efforts within the community. 

The results revealed significant differences in measuring efficiency between the two groups, addressing Research Question 1 (RQ1). The intervention group, which used the game-based app, demonstrated superior performance compared to the control group, which utilized the conventional non-game app.

The game-based app group exhibited significantly better efficiency retention than the control group, addressing Research Question 2 (RQ2). This suggests that while both groups showed improvements in measuring efficiency, the game-based app not only enhanced initial performance but also maintained higher levels of efficiency over time. These findings underscore the potential benefits of integrating gamification into data collection processes, enhancing both immediate outcomes and long-term retention of performance improvements.

\section{Future Work}

While this study effectively compares the effectiveness of serious games in anthropometric measurement of children, it does not address the long-term outcomes of the intervention due to constraints related to resources and time. Future research should aim to explore how participants' approaches and engagement evolve over extended periods in both the game-based and traditional data collection conditions. Understanding these dynamics can provide insights into the sustained impact of gamification on motivation and performance. Future studies could investigate the underlying factors contributing to observed differences in motivation and effectiveness, potentially leading to the development of more refined and long-lasting interventions. Expanding research to include longitudinal evaluations and qualitative assessments will enhance our understanding of how serious games influence long-term engagement and outcomes in community health settings.

\begin{acks}
This study was jointly funded by the Science and Engineering Research Board (SERB), the Federation of Indian Chambers of Commerce and Industry (FICCI), and UNICEF India. We extend our sincere gratitude to all the Community Healthcare Workers and their supervisors who actively participated in this research. Their invaluable contributions were essential to the success of this study. We appreciate the constructive feedback provided by the anonymous reviewers, which greatly enhanced the quality and depth of our work.
\end{acks}

\bibliographystyle{ACM-Reference-Format}
\bibliography{references}

@inproceedings{Birk2017,
    title = {{Age-based preferences and player experience: A crowdsourced cross-sectional study}},
    year = {2017},
    booktitle = {CHI PLAY 2017 - Proceedings of the Annual Symposium on Computer-Human Interaction in Play},
    author = {Birk, Max V. and Friehs, Maximillian A. and Mandryk, Regan L.},
    month = {10},
    pages = {157--170},
    publisher = {Association for Computing Machinery, Inc},
    isbn = {9781450348980},
    doi = {10.1145/3116595.3116608}
}

@incollection{Deterding2018,
    title = {{An Introduction to the Gameful World}},
    year = {2018},
    booktitle = {The Gameful World},
    author = {Walz, Steffen P. and Deterding, Sebastian},
    doi = {10.7551/mitpress/9788.003.0001}
}

@article{Mishra2024,
    title = {{Association between the use of Accredited Social Health Activist (ASHA) services and uptake of institutional deliveries in India}},
    year = {2024},
    journal = {PLOS Global Public Health},
    author = {Mishra, Sujata and Horton, Susan and Bhutta, Zulfiqar A. and Essue, Beverley M.},
    editor = {Gaurav, Sarthak},
    number = {1},
    month = {1},
    pages = {e0002651},
    volume = {4},
    url = {https://dx.plos.org/10.1371/journal.pgph.0002651},
    doi = {10.1371/journal.pgph.0002651},
    issn = {2767-3375}
}

@inproceedings{Kappen2016,
    title = {{Design strategies for gamified physical activity applications for older adults}},
    year = {2016},
    booktitle = {Proceedings of the Annual Hawaii International Conference on System Sciences},
    author = {Kappen, Dennis L. and Nacke, Lennart E. and Gerling, Kathrin M. and Tsotsos, Lia E.},
    month = {3},
    pages = {1309--1318},
    volume = {2016-March},
    publisher = {IEEE Computer Society},
    isbn = {9780769556703},
    doi = {10.1109/HICSS.2016.166},
    issn = {15301605}
}

@inproceedings{Erdfelder2007,
    title = {{G*Power 3: A flexible statistical power analysis program for the social, behavioral, and biomedical sciences}},
    year = {2007},
    booktitle = {Behavior Research Methods},
    author = {Faul, Franz and Erdfelder, Edgar and Lang, Albert Georg and Buchner, Axel},
    number = {2},
    volume = {39},
    doi = {10.3758/BF03193146},
    issn = {1554351X}
}

@article{Asthana2022,
    title = {{India's one million Accredited Social Health Activists (ASHA) win the Global Health Leaders award at the 75th World Health Assembly: Time to move beyond rhetoric to action?}},
    year = {2022},
    journal = {The Lancet Regional Health - Southeast Asia},
    author = {Asthana, Sumegha and Mayra, Kaveri},
    pages = {100029},
    volume = {3},
    url = {https://doi.org/10.1016/j.},
    doi = {10.1016/j}
}

@misc{Heidkamp2021,
    title = {{Mobilising evidence, data, and resources to achieve global maternal and child undernutrition targets and the Sustainable Development Goals: an agenda for action}},
    year = {2021},
    booktitle = {The Lancet},
    author = {Heidkamp, Rebecca A. and Piwoz, Ellen and Gillespie, Stuart and Keats, Emily C. and D'Alimonte, Mary R. and Menon, Purnima and Das, Jai K. and Flory, Augustin and Clift, Jack W. and Ruel, Marie T. and Vosti, Stephen and Akuoku, Jonathan Kweku and Bhutta, Zulfiqar A.},
    number = {10282},
    volume = {397},
    doi = {10.1016/S0140-6736(21)00568-7},
    issn = {1474547X}
}

@misc{Bochennek2007,
    title = {{More than mere games: A review of card and board games for medical education}},
    year = {2007},
    booktitle = {Medical Teacher},
    author = {Bochennek, Konrad and Wittekindt, Boris and Zimmermann, Stefanie Yvonne and Klingebiel, Thomas},
    number = {9-10},
    volume = {29},
    doi = {10.1080/01421590701749813},
    issn = {0142159X}
}

@article{NFHS5IndiaFactsheet,
    title = {{National Family Health Survey-5 India Fact Sheet}},
    year = {2021},
    journal = {Indian Institute Population Sciences and Ministry of Health and Family Welfare},
    author = {{Indian Institute Population Sciences and Ministry of Health and Family Welfare}},
    url = {http://rchiips.org/nfhs/factsheet_NFHS-5.shtml}
}

@inproceedings{Majhi2022,
    title = {{Physical and Augmented Reality based Playful Activities for Refresher Training of ASHA Workers in India}},
    year = {2022},
    booktitle = {Conference on Human Factors in Computing Systems - Proceedings},
    author = {Majhi, Arka and Agnihotri, Satish and Mondal, Aparajita},
    doi = {10.1145/3516492.3558788}
}

@article{Majhi2024c,
    title = {{Refresher Training through Digital and Physical, Card-Based Game for Accredited Social Health Activists (ASHAs) and Anganwadi Workers (AWWs) in India}},
    year = {2024},
    journal = {Companion Proceedings of the Annual Symposium on Computer-Human Interaction in Play (CHI PLAY Companion '24), October 14-17, 2024, Tampere, Finland},
    author = {Majhi, Arka and Mondal, Aparajita and Agnihotri, Satish B},
    volume = {1},
    publisher = {ACM},
    url = {https://doi.org/10.1145/3665463.3678819},
    doi = {10.1145/3665463.3678819}
}

@inproceedings{Majhi2021,
    title = {{Refresher Training through Quiz App for capacity building of Community Healthcare Workers or Anganwadi Workers in India}},
    year = {2021},
    booktitle = {5th Asian CHI Symposium 2021},
    author = {Majhi, Arka and Joshi, Anirudha and Agnihotri, Satish B. and Mondal, Aparajita},
    doi = {10.1145/3429360.3468186}
}

@incollection{Majhi2024a,
    title = {{Replay, Revise, and Refresh: Smartphone-Based Refresher Training for Community Healthcare Workers in India}},
    year = {2024},
    author = {Majhi, Arka and Mondal, Aparajita and Agnihotri, Satish B.},
    pages = {310--320},
    url = {https://link.springer.com/10.1007/978-3-031-61966-3_34},
    doi = {10.1007/978-3-031-61966-3{\_}34}
}

@inproceedings{Zimmerman2007,
    title = {{Research through design as a method for interaction design research in HCI}},
    year = {2007},
    booktitle = {Conference on Human Factors in Computing Systems - Proceedings},
    author = {Zimmerman, John and Forlizzi, Jodi and Evenson, Shelley},
    doi = {10.1145/1240624.1240704}
}

@inproceedings{Shah2017,
    title = {{Tackling child malnutrition: An innovative approach for training health workers using ICT: A pilot study}},
    year = {2017},
    booktitle = {IEEE Region 10 Humanitarian Technology Conference 2016, R10-HTC 2016 - Proceedings},
    author = {Shah, Mithilesh P. and Kamble, Pawan A. and Agnihotri, Satish B.},
    isbn = {9781509041770},
    doi = {10.1109/R10-HTC.2016.7906811}
}

@misc{JuhoHamari2019,
    title = {{The rise of motivational information systems: A review of gamification research}},
    year = {2019},
    booktitle = {International Journal of Information Management},
    author = {Koivisto, Jonna and Hamari, Juho},
    volume = {45},
    doi = {10.1016/j.ijinfomgt.2018.10.013},
    issn = {02684012}
}

@article{Harris2006,
    title = {{The use and interpretation of quasi-experimental studies in medical informatics}},
    year = {2006},
    journal = {Journal of the American Medical Informatics Association},
    author = {Harris, Anthony D. and {et.al.}},
    number = {1},
    month = {1},
    pages = {16--23},
    volume = {13},
    doi = {10.1197/jamia.M1749},
    issn = {10675027},
    pmid = {16221933}
}

@article{Jordan2013,
    title = {{Turning life into a game: Foursquare, gamification, and personal mobility}},
    year = {2013},
    journal = {Mobile Media and Communication},
    author = {Frith, Jordan},
    number = {2},
    volume = {1},
    doi = {10.1177/2050157912474811},
    issn = {20501587}
}

@techreport{DOH2013,
    title = {{WMA Declaration of Helsinki - Ethical Principles for Medical Research Involving Human Subjects}},
    year = {2013},
    author = {{World Medical Association}},
    url = {https://www.med.or.jp/dl-med/wma/helsinki2013e.pdf}
}

\end{document}